
\documentstyle{amsppt}
\nopagenumbers
\nologo
\hsize 32pc
\vsize 50pc
\emergencystretch=100pt

\def\ms{{\medskip}}

\def\k{{\kappa}}
\def\t{{\tau}}

\def\slr{{\Cal R}}

\def\slp{{\Cal P}}

\def\smalltype{\let\rm=\eightrm \let\bf=\eightbf
\let\it=\eightit \let\sl=\eightsl \let\mus=\eightmus
\baselineskip=9.5pt minus .75pt \rm}
\parindent=30pt

\def\onepsi{\hbox{\hbox{$^1$}\kern-.15em $\Psi$}}
\def\onejt{\hbox{\hbox{$^1$}\kern-.25em $\tilde J$}}
\def\onext{\hbox{\hbox{$^1$}\kern-.25em $\tilde X$}}
\def\oneit{\hbox{\hbox{$^1$}\kern-.25em $\tilde I$}}
\def\onert{\hbox{\hbox{$^1$}\kern-.1em $\tilde \slr$}}
\def\oneslr{\hbox{\hbox{$^1$}\kern-.1em $\slr$}}

\def\udots{\mathinner{\mkern1mu\raise1pt\vbox{\kern7pt\hbox{.}}\mkern2mu
\raise4pt\hbox{.}\mkern2mu\raise7pt\hbox{.}\mkern1mu}}
\def\uudots{\mathinner{\mkern1mu\raise2pt\vbox{\kern7pt\hbox{.}}\mkern2mu
\raise5pt\hbox{.}\mkern2mu\raise8pt\hbox{.}\mkern1mu}}

\magnification =1200
\def\myprime{^\prime}
\pagewidth{ 12 cm }
\pageheight{ 7.5 in}
\baselineskip=25pt plus 2pt
\font\twelverm=cmr12
%
%
\topmatter

\title
{Localized induction equation}
{and}
{pseudospherical surfaces}
\endtitle
\author
Ron Perline
\endauthor
\affil
Dept. of Mathematics and Computer Science, Drexel University
\endaffil
\abstract
We describe a close connection between the
localized induction equation
hierarchy of integrable evolution equations on
space curves, and surfaces of constant negative
Gauss curvature.
 $$ \quad $$
 {\twelverm
 \centerline{ To appear in}
 \centerline{}
 \centerline{Journal of Physics A:  Mathematical
 and General} }
 $$\quad $$
$$\quad $$
\endabstract
\endtopmatter

\newpage

{\bf 1. Introduction}.
\medskip
Many of the integrable equations of non-linear science
have essentially equivalent realizations in terms of the
classical geometry of curves and surfaces in space.  These
geometric realizations provide new insight into the structure
of the integrable equations; in addition, these geometric
problems may well have interesting physical interpretations
in their own right.  In this paper, we describe recent
developments illustrating a close
connection between two such geometric realizations: the
localized induction equation (LIE) and pseudospherical
surfaces, or surfaces of constant negative
Gauss curvature.
\newline
(1) {\it Localized induction equation}:
LIE is a local geometric evolution equation defined
on space curves via the equation
$$\gamma_t = \gamma_s \times \gamma_{ss}, $$
where $s$ is the arclength parameter for the evolving
space curve $\gamma(s,t) \in R^3$, and $\times$ denotes
cross product.  When the curvature is non-vanishing, the
right-hand side can be written $\kappa B$, where $B$
is the binormal to $\gamma$, and $\kappa$ is the curvature.
LIE was developed in fluid mechanics as an idealized
{\it local} model for the evolution
of the centerline  of a thin, isolated
vortex tube in an inviscid fluid (for derivation and history,
see [1],[2],[3]; for a discussion of more accurate, non-local
models, see [4],[5]). As in the case of the full
inviscid Euler
equations from which it is derived, LIE can be described
as a Hamiltonian evolution equation, and in fact
the corresponding Hamiltonian is just the length functional
on space curves [6].
The connection of LIE to soliton theory was  made apparent through a
discovery  of Hasimoto [7]:  if $\gamma$ evolves according
to LIE, then the induced evolution of its {\it complex curvature}
$\psi = \kappa e^{i \int^s \tau(u) \, du}$
($\tau$ is torsion along the curve)
is given by the {\it cubic non-linear Schr\"odinger equation} (NLS)
$\psi_t = i(\psi_{ss} + {1 \over 2}|\psi|^2 \psi)$. NLS is a
well-known example of a completely integrable evolution equation;
the result of Hasimoto implies that LIE is a geometric realization
for NLS.
{}Further investigations of the LIE-NLS correspondence were reported
in [8],[9] and details of the
complete integrability of LIE itself are also described there.
We remark that the connection between
equations of NLS-type
and the equations of fluid motion
remains a topic of current research
[10].
\newline
(2) {\it Pseudospherical surfaces}:
The study of pseudospherical surfaces in Euclidean space spans a period of
more than a  century; in particular, we
mention the early works  of Dobriner, Enneper,
and B\"acklund [11],[12],[13].
Recent interest has been spurred by the connection to soliton
theory [14],[15] (a kindred problem, finding metrics on $R^2$
with constant curvature, is also related to integrable
evolution equations: see [16],[17]).
 We mention two such connections:
\newline
(i) Given a pseudospherical surface $M$,
the angle $\psi$ between its asymptotic curves
satisfies the
{\it sine-Gordon equation} (SG)  ${{\partial^2 \psi} \over {\partial x y}} =
sin (\psi)$, where $x$ and $y$ are asymptotic coordinates for the
surface (for basic  definitions from surface theory, see [18]).
Again, SG is a well-known example of a completely integrable equation,
which arises in numerous physical problems [19],[20],[21].  Thus, a
pseudospherical surface $M$ is a geometric realization of a given solution
to SG.
\newline
(ii)  Given a pseudospherical surface $M$,  its second fundamental form induces
a Lorentz metric on the surface.  The Gauss map of $M$ (taking $M$ to
the two sphere $S^2$) is a {\it harmonic map} [14],[15].
This is an example of a classical chiral model, for which there exists
an extensive literature ([22],[23],[24],[25] and references therein).
\newline

We now make a simple observation  which demonstrates
that LIE  has {\it some} connection to surface theory. Consider
any curve $\gamma = \gamma(s,0)$ and let it evolve according
to LIE.
Because $N$ is normal to the resulting swept-out surface, it follows that
$\gamma(s,t)$ is a geodesic for any time $t$,
thus providing a
geodesic foliation of the resulting surface.

To describe the connection with pseudospherical surfaces, we make
reference to  the complete integrability properties of LIE.  LIE is
the first (nontrivial) term of an infinite sequence of commuting Hamiltonian
evolution  equations on curves, all of which  equations are local-geometric
in nature; we call this sequence the {\it localized
induction hierarchy} (LIH).
The associated Hamiltonians (which are conserved quantities
for LIE) can be expressed as global geometric invariants of the curves.
We shall see that certain distinguished {\it soliton curves}
(= critical points for linear combinations of the Hamiltonians), after
evolving according to a  related linear combination  of evolution equations
from LIH, sweep out pseudospherical surfaces.
In analogy with the geodesic construction
of the previous paragraph, the induced foliation plays a role
in the geometry of the surface: the curves of the foliation
are asymptotic lines for the surface. The main point of this paper
is to describe this construction.
We also find an interesting connection between pseudospherical surfaces
and B\"acklund transformations for certain curves; see Section 4.
In this same section, there is a surprising technical result
suggesting deeper relations to Lie groups:  two natural
bases for  a geometrically defined vector space, relevant to our
theory, are related via a change-of-basis matrix defined in
terms of lower triangular Toeplitz matrices.
In the last section we discuss
a related topic: evolution equations on surfaces which preserve
the pseudosphericity property.
For brevity, proofs have been  omitted, but sufficient computational detail
is presented so that the reader can at least reconstruct the basic
examples described here.

One way of viewing  our technique
is as a    ``nonlinear factorization" of the problem  of constructing
pseudospherical surfaces: the simpler ``factors" are the related
variational problem
on curves, and then the subsequent evolution of critical points of this
variational problem according to appropriate evolution equations.
Historically, we know that solution techniques for integrable systems
``travel well": if applicable to one integrable example, they can
usually be modified to apply to essentially all other known integrable
problems.  Thus, this study of the LIE-pseudospherical
connection will hopefully
have consequences for the study of integrable models of more direct
interest to mathematical physics.

 %
 %
 %

\medskip
\medskip
{\bf 2. LIH and related hierarchies}
\medskip
As stated above, LIE belongs to an infinite hierarchy of evolution
equations on curves, all of the form $\gamma_t = X_n = aT + bN + cB$, where
$\{ T,N,B \}$ is the Frenet frame along the curve,
and $a,b,c$ are functions (polynomial) of   $\kappa, \tau, \kappa' =
\kappa_s, \tau' = \tau_s$, and higher derivatives with respect to $s$.
We list the first
few terms of the hierarchy, as well as their associated
Hamiltonians (the vector field $X_0$ is exceptional):
$$
\eqalign{
& X_{0} = -T,   \cr
& X_{1} = \kappa B, \quad  I_{1} = \int_\gamma \ ds , \cr
& X_{2} = {{\kappa^2} \over 2 }T  +   \kappa \myprime N + \kappa \tau B,
 \quad I_{2} = \int_\gamma -\tau \ ds ,\cr
& X_{3} \ =
\kappa^2 \tau T
+ (2 \kappa\myprime \tau + \kappa \tau \myprime)N
+ (\kappa \tau^2 - \kappa ^{\prime \prime} -{1 \over 2}\kappa^3)B ,  \cr
& \quad I_3 = \int_\gamma {1 \over 2} \kappa^2 \ ds , \cr
& X_{4} \ =
(-\k \k '' + {1 \over 2} (\k ')^2 + {3 \over 2} \k ^2 \t ^2 - {3 \over 8} \k
^4)T \cr
& + ( - \k ''' + 3 \k \t \t ' + 3 \k ' \t ^2 - {3 \over 2} \k ^2 \k ') N \cr
& + ( \k \t ^3 - 3 (\k ' \t ) ' -  {3 \over 2} \k ^3 \t - \k \t '') B, \cr
& \quad I_{4} = \int_\gamma  \, {1 \over 2} \kappa^2 \tau \,{ds}, \cr
& \dots \cr
}
$$
The vector fields of LIH are {\it locally arclength preserving}
(LAP): a vector field $W$ is LAP if every segment of a curve $\gamma$
has its length remain constant as $\gamma$ evolves via $\gamma_t = W$.
Equivalently, $<W_s,T> = 0$.

The first few functionals in the list have simple physical
interpretation.
As shown in [26], critical points of linear combinations of
the functionals $I_1,I_2,I_3$
are the Kirchhoff rods of elasticity theory.
Interestingly, these are exactly the curves whose shape remains
unchanged as they evolve according to LIE ([9],[27],[28]).
Another discussion of the physical interpretation of the invariants
of LIE can be found in [29].

As is usually the case with integrable systems, LIH is generated
by a recursion operator $X_{n+1} = \slr X_n, \ n \ge 0$; if
$X = aT + bN + cB$ then $\slr(X) = -\slp(T \times X')$, where
$\slp$ is a {\it parameterization operator}
$\slp(X) = {\int^s (\kappa b) ds}\, T + bN + cB$. Besides being
useful for generating  LIH, $\slr$ can be used to compactly express
the first-order variations in curvature and
torsion along {\it any} vector field
$W$ which is LAP [9]:
$$W(\kappa) = <-\slr^2(W),N>,$$
$$W(\tau) = < -\slr^2(W),B/\kappa >' \, .$$
Related formulas
also exist for the evolution of frame fields along
$W$ [30].

There are a number of hierarchies of integrable geometric evolution
equations, related to LIE,  which have interesting geometric properties.
These are discussed in more detail in [31]; we mention those which
are relevant here:
\newline
(1) {\it Constant torsion preserving (CTP)}:  For $n \ge 0$, the
vector fields
$$Z_n = \sum_{k=0}^{2n} {{\binom {2n+1}k} (-\tau_0)^k X_{2n-k}}$$
preserve the constant torsion condition $\tau = \tau_0$.
If a constant torsion curve $\gamma$ evolves according to
$\gamma_t = Z_n$, the induced evolution on curvature
$\kappa_t = Z_n(\kappa)$ is the corresponding element
of the (mKdV) hierarchy; in particular, $Z_1$ induces the
(mKdV) evolution $\kappa_t = \kappa_{sss} + {3 \over 2} \kappa^2 \kappa_s$,
recovering a result of Lamb [32].
Recently, Fukumoto and Miyazaki [33]
have derived a refined version of LIE
which allows for axial velocity for the vortex tube: modulo trivial scaling
terms, their equation is exactly $\gamma_t = Z_1$.
\newline
(2) {\it  Planar preserving}: A special case of (i) is worth remarking;
when $\tau_0 = 0$, the sequence $Z_n$ just reduces to the even
$X_{2n}$ restricted to planar curves.  This integrable hierarchy
of evolution equations has been discussed by several authors
[34],[35],[36]. The first term of the hierarchy can
be interpreted physically:
in [37], it is shown that $\gamma_t = Z_1$,
when restricted to planar curves, is a ``localized induction equation"
for  boundary curves of vortex patches for 2D ideal fluid flow.
The {\it even} functionals $I_{2n}$ for LIH vanish identically on
planar curves, and the {\it odd} functionals $I_{2n+1}$ restrict
to give functionals on planar curves which depend only on $\kappa$
and its derivatives.
\newline
(3) {\it Torsion independent}: The
vector fields $A_0 = -T$,
$$A_n = \sum_{k=0}^{n-1} {{\binom {n-1}k} (-\tau_0)^k X_{n-k}} ,
\quad n \ge 1$$
have the property that, along curves $\gamma$ with $\tau = \tau_0$, the
coefficients of $A_n = aT + bN + cB$ have no explicit $\tau$ dependence.
The odd vector fields in the sequence are purely binormal; the even
vector fields, on the other hand, have $0$ binormal component. We thus
refer to the even fields as ``planar-like" and introduce the notation
$\Omega_n = A_{2n}$.
\newline
\newline
{\bf 3. Pseudospherical surfaces and the ``trigonometric equation"}
\medskip
We briefly review basic facts from surface theory in $R^3$, mostly
to establish notation and terminology.
Given an oriented surface $M$, the {\it Gauss
map} $\nu: M \rightarrow S^2$ sends a point $p \in M$ to its unit normal.
By identifying tangent spaces $T_pM$ and $ T_{\nu(p)}S^2$, one obtains
the {\it Weingarten map} $\ -d\nu: T_pM \rightarrow T_pM$.
The {\it second fundamental form} is given by $\Pi(w) = <-d\nu(w),(w)>$, for
any $w \in T_pM$.
The determinant
of the Weingarten map is the {\it Gauss curvature} of $M$.  If the Gauss
curvature  is negative, then at any point $p$ there will be two linearly
independent vectors $v_i, \,  i=1,2$ such that $\Pi(v_i) = 0$: these are
the {\it asymptotic directions} of the surface.  Any curve whose tangent
at every point corresponds to an asymptotic direction is called an
{\it asymptotic curve} or {\it line}.
If $M$ is pseudospherical, then $M$ has two transverse
foliations by asymptotic lines. A theorem of Beltrami-Enneper [38]
states that the Gauss curvature of a surface $M$ along an asymptotic
line $\gamma$ is the negative of the square of the torsion $\tau$ of
$\gamma$; if $M$ is pseudospherical, then its
asymptotic lines have {\it constant}
torsion.

Conversely, given a curve $\gamma$ with constant torsion $\tau_0$,
there is a dynamical prescription for finding a
pseudospherical surface $M$ with
$\gamma$ as an asymptotic line:

{\it Proposition: Let $\gamma = \gamma(s,0)$ be the initial condition
for the ``trigonometric equation" $\gamma_t = W =   cos(\theta)T -
sin(\theta)N$,
where $\theta = \int^s \kappa(u) du$.
The resulting swept-out surface $M$ is pseudospherical with
curvature $G= -\tau_0^2$.  For any $t$, $\gamma(s,t)$ is an asymptotic
curve for $M$.  The induced evolution of  $\theta$ is given
by the sine-Gordon equation $\theta_{st} = -G sin(\theta)$.}
\newline
{}For a discussion and proof, see [31],[39].
\newline
\newline
{\bf 4.  Planar-like solitons and pseudospherical surfaces}
\medskip
(1) {\it Planar and planar-like solitons}: As stated above, the
odd functionals for LIE restrict to planar curves. Let
$J_n$ denote the restriction of $I_{2n+1}$ to planar curves;  such
functionals depend upon curvature only.  A {\it planar soliton} is
a planar  curve which is a critical point for a linear combination of the
$J_n$.  For example, critical points for $J_1 + aJ_0 =
\int_\gamma ({{1 \over 2}\kappa^2 + a})\, ds$ have curvature
functions  satisfying the Euler-Lagrange equation
$$\kappa{''}  +{1 \over 2}\kappa^3 - a\kappa =0 . $$
$J_1$ represents elastic energy for a curve, $J_0$ a length constraint;
the associated critical points are called {\it planar elastic curves}
or {\it elastica}.

For simplicity, we specify boundary conditions of {\it asymptotic linearity}
on our curves by assuming that
$\kappa$ and its derivatives
vanish
as $s \rightarrow
\pm \infty$.  For each $J_n$, we denote its associated Euler operator
by $E_n$; the first three are:
$$\eqalign{
E_0(\kappa) &= -\kappa(s), \cr
E_1(\kappa) &= {\frac {d^{2}}{ds^{2}}}\kappa(s)+{\frac {\kappa(s)^{3}}{2}}, \cr
E_2(\kappa) &= -{\frac {d^{4}}{ds^{4}}}\kappa(s)
-{\frac {5\,\kappa(s){\frac {d}{ds}}
\kappa(s)^{2}}{2}}-{\frac {5\,\kappa(s)^{2}{\frac {d^{2}}{ds^{2}}}
\kappa(s)}{2}}-{\frac {3\,\kappa(s)^{5}}{8}} \cr
}
$$

A {\it planar-like soliton} is a space curve $\gamma$ which constant
torsion $\tau = \tau_0$ whose curvature $\kappa$ is the same as that
of a planar soliton.
Thus, $\sum_{i=0}^n a_i E_i(\kappa) = 0$ for some choice of constants
$a_i$. This shows that planar-like solitons are {\it related} to
critical points of the geometric functionals associated to LIE; the
next proposition states that they {\it are} critical points for
appropriate functionals:
\medskip
{\it
Proposition:  Let a space curve $\gamma$ be a planar-like soliton
with torsion $\tau_0$ and curvature satisfying
$$\sum_{i=0}^n a_i E_i(\kappa) = 0 . $$  Then $\gamma$ is a critical
point of the functional
$$\sum_{i=0}^n a_i (\sum_{j=0}^{2i}{
{\binom{2i}j}(-\tau_0)^j I_{2i+1 -j})
};
$$
Equivalently, the vectorfield $\sum_0^n a_i A_{2i+1}$ vanishes along $\gamma$.
}
\newline
In the last proposition, the planar-like solitons are distinguished
critical points in that they have constant torsion, which will not
be true in general.
\newline
(2) {\it s-integrals }:
The $E_i$ previously mentioned can be used to construct the mKdV hierarchy
of integrable evolution equations via $\kappa_t =
 {d \over ds}E_i(\kappa), \, i \ge 0$.
It is a part of the general theory of these equations [40] that,
associated to the Euler operators
$E(\kappa) = \sum_{i=0}^n a_i E_i(\kappa) $ are
{\it s-integrals} $T_j(\kappa)$, where
$E(\kappa){ d \over {ds}}E_j(\kappa) = {d \over {ds}}T_j(\kappa),
\, j=0,1, \dots n-1$.  The
$T_j$ are polynomial expressions in $\kappa$ and its derivatives.
Along a planar-like soliton,
we have $E(\kappa)=0$, so $T_j = c_j$.  In fact, for asymptotically
linear curves, the $c_j$ are
all $0$.
\newline
(3) {\it Definition and properties  of $T^*$}: For the rest of this section,
$\gamma$ will refer to a planar-like
$n$-soliton
with torsion $\tau_0 \ne 0$ and curvature $\kappa$ satisfying
$ E(\kappa) = \sum_{i=0}^n a_i {\tau_0}^{-2n} E_i(\kappa) =
0 $ with $a_0 \ne 0$,
$a_n \ne 0$;
we set $b_i = a_i {\tau_0}^{-2n}$.
We define a planar-like vector field along $\gamma$  (no binormal component)
$$T^* = (-1/b_0)(\sum_{i=0}^n b_i \Omega_i) \, .$$ We describe the
properties  of the evolution equation
$\gamma_t = T^*$ with  our planar-like soliton $\gamma$ as its
initial condition - we call the reader's attention in particular
 to articles  (iv) and (vii):
\newline
(i) {\it $T^*$ is CTP along $\gamma$}:
To see this, one proves
the identity
$$  T^* = {{-1} \over {b_0 \tau_0^2}}{\sum_{k=0}^{n+1} (b_{k-1} -2b_k\tau_0^2
+ b_{k+1}\tau_0^4)Z_k} ,$$
thus expressing $T^*$ in terms of the  CTP vectorfields $Z_n$.
The variation in curvature associated with the evolution
$\gamma_t = T^*$ is given by
$$  \kappa_t =   {{-\tau_0^2 } \over {b_0}}
\sum_{k=0}^{n-1}{b_{k+1} {d \over {ds}}E_k} , $$
a combination of terms in the mKdV hierarchy.
\newline
(ii) {\it $T^*$ preserves soliton type}: As indicated above,
$\gamma$ is a critical point
for a linear combination of conserved functionals for the LIE hierarchy,
distinguished by
having constant torsion.  Since $T^*$ itself is a
linear combination of terms from LIE, it
deforms $\gamma$ into another critical point; (i) shows
that the deformation preserves
constant torsion.
\newline
(iii) {\it $T^*$ is of unit length along $\gamma$}:
This is a direct consequence of $E(\kappa) = 0$.
\newline
(iv) {\it Geometry of the swept-out surface}:
Let $M$ be the surface swept out via the
evolution $\gamma_t =T^*$.  For any time $t$,
$T^*$ is a linear combination of the Frenet
vectors $T$ and $N$; hence the normal $\nu$ to the
surface $M$ is $B$, the binormal
to the curve $\gamma$.  We compute $\Pi(T) = <-d\nu(T),T >
= <-\nabla_T B, T> = < \tau N, T> = 0$;
$T$ is an asymptotic direction for $M$.
By the Beltrami-Enneper theorem,
{\it $M$ is a pseudospherical surface}
with Gauss curvature $G = -\tau_0^2$. We will call a
pseudospherical surface $M$ a {\it soliton surface} if its
asymptotic curves consist of planar-like solitons.
\newline
(v) {\it $T^*$ as an asymptotic direction}:
To show that $T^*$ is another asymptotic direction
for $M$, one needs to compute
$\Pi(T^*) = <-d\nu(T^*),T^* > = <-\nabla_{T^*} B, T^*>$. The term
$\nabla_{T^*} B$ requires the variation formulas for
frames  derived in [30] which  were  mentioned above; the result
is that $T^*$ is indeed an asymptotic direction.
We call $T^*$ the {\it conjugate} asymptotic
direction and its integral curve $\gamma^*$ the
{\it conjugate} asymptotic curve.  By (iii),
$T^*$ is the unit tangent vector along $\gamma^*$.
Since $M$ is pseudospherical, it must be the case
that the torsion of $\gamma^*$ is $\pm \tau_0$;
a calculation shows it to be $\tau_0$.
\newline
(vi) {\it $\kappa^*$ in terms of $\kappa$}:  At a point $p$ on $M$,
the conjugate curvature of
$\gamma^*$ can be expressed in
terms of the curvature of $\gamma$ at that point:
$$  \kappa^* = {{-\tau_0^2} \over {b_0}}{\sum_0^{n-1} b_{i+1} E_i(\kappa)},$$
Again, the  frame variation formulas of [30] are used to
derive the Frenet equations
for $\gamma^*$ and hence $\kappa^*$.
\newline
(vii) {\it $\gamma^*$ is a planar-like soliton}:  by (v),
 we know that $\gamma^*$ has
torsion $\tau_0$. The curvature function satisfies the equation
$$ E^*(\kappa^*) = \sum_{i=0}^n b_{i}^* E_i^*(\kappa^*) = 0, $$
where $E_i^*$ denotes the Euler operator $E_i$,
with differentiation with respect to
$s$ replaced by differentiation with
respect to $s^*$ ($=t=$ arclength along $\gamma^*$);
$b_i^* = a_i^* \tau_0^{-2n}$, where $a_i^* = a_{n-i}$.
{\it $\gamma^*$ is therefore
a planar-like soliton of the same order as $\gamma$,
with ``flipped" coefficients}.
The existence of a pseudospherical surface
containing both $\gamma$ and $\gamma^*$ as asymptotic
lines provides a {\it geometric B\"acklund transformation for
planar-like solitons}.
The proof
requires use of  the $s$-constants of motion of section (2),
and the variation of curvature
described in article (i).
\newline
(viii) {\it $T^*$ and the ``trigonometric equation"}:
We have already seen the connection
between curves of constant torsion, pseudospherical surfaces,
and the unit-length
vectorfield $W = cos(\theta)T -
sin(\theta)N$. Let $A = cos(\theta), B= -sin(\theta)$.
Then $A,B$ satisfy the differential
equations ${d \over {ds}}A = \kappa(s)B$, ${d \over {ds}}B = -\kappa(s)A$.
 $T^*$  is also
unit length; and $T^* = F T + G N$, where $F,G$ are
polynomial expressions in $\kappa$ and
its derivatives.  One can check that along $\gamma$,
 $F,G$ satisfy the same differential equations
as do $A,B$; and at $s = -\infty$, their respective values agree.
 This shows that along
planar-like solitons, the ``trigonometric" vectorfield can be
expressed in terms of {\it local}
quantities associated with the curve.
\newline
(ix) {\it The conjugate LIE hierarchy}:
 By definition, the vectorfield $T^*$ can be expressed
as a linear combination of the vectorfields $X_n$.
One can think of $T^*$ as (minus) the zeroth
term in the {\it conjugate} LIE hierarchy and ask if the
higher order terms are also expressible
in terms of the LIE hierarchy along $\gamma$.
 By the conjugate hierarchy we mean
vectorfields such as $X_1^* = \kappa^*B^* = \kappa^*B$, and so forth.

It is actually more convenient to  express the relation between the
vectorfields
$A_n^*$ and $A_n$; this is essentially equivalent
information since along a constant torsion
curve the  $A_n$ span the same space as the $X_n$.
Also, along  planar-like solitons, we have   $\sum_0^n b_i A_{2i+1} =0$,
so we need
only consider the span of $A_1, \dots , A_{2n}$
(an analogous statement holds for $\gamma^*$).

{\it Proposition: Along a planar-like soliton $\gamma$,
the $n$ vectorfields $A_{2i-1}^*, i = 1, \dots ,n$ can be
expressed as a linear combination
of the vectorfields $A_{2i-1}, i = 1, \dots, n$, and a
similar statement holds for the $A_{2i},
i=1,\dots,n$.
In particular
$$A_{2i-1}^* = \sum_{j=1}^n ({\Cal S}^{-1}{\Cal H}
{\Cal T})_{ij} A_{2j-1} \, ,$$
where $\Cal S$ is the Toeplitz matrix
$$
{\Cal S} = \pmatrix
\format\quad\c&\quad\c&\quad\c&\quad\c\quad\\
b_0&0&\hdots&0\\
b_1&\ddots&&\vdots\\
\vdots&\ddots&\ddots&0\\
b_{n-1}&\hdots&b_1&b_0\\
\endpmatrix \ ,
$$

${\Cal H}$  is the Hankel matrix
$$
{\Cal H} = \pmatrix
\format\quad\c&\quad\c&\quad\c&\quad\c\quad\\
0&\hdots&0&1\\
\vdots&&\uudots&0\\
0&\udots&&\vdots\\
1&0&\hdots&0\\
\endpmatrix \ ,
$$
and $\Cal T$ is the Toeplitz matrix
$$
{\Cal T} = \pmatrix
\format\quad\c&\quad\c&\quad\c&\quad\c&\quad\c\quad\\
b_n&0&\hdots&0\\
b_{n-1}&\ddots&&\vdots\\
\vdots&\ddots&\ddots&0\\
b_1&\hdots&b_{n-1}&b_n\\
\endpmatrix \ .
$$
A similar transformation  exists relating the $A_{2i}^*$ and $A_{2i}$:
it is given
by ${\Cal S}^{-1} {\Cal K}{\Cal T}$, where ${\Cal K}$
is  the ``almost Hankel" matrix
$$
{\Cal K} = \pmatrix
\format\quad\c&\quad\c&\quad\c&\quad\c&\quad\c&\quad\c\quad\\
0&0&\hdots&0&1&{-b_1/b_0}\\
\vdots&\vdots&0&1&0&\vdots\\
\vdots&\udots&\udots&\udots&&\vdots\\
0&1&0&\hdots&0&{-b_{n-2}/b_0}\\
1&0&\hdots&\hdots&0&{-b_{n-1}/b_0}\\
0&\hdots&\hdots&\hdots&0&{-b_n/b_0}\\
\endpmatrix \ .
$$
}
This proposition is relevant to the discussion in Section 5.
\newline
(x) {\it Symmetry}: For a pseudospherical surface $M$, we have been discussing
an asymptotic curve
$\gamma$ and its conjugate curve $\gamma^*$.
Of course, there is a symmetric relation
between these two curves:  $\gamma$ can be thought of
as the conjugate curve for $\gamma^*$.
The formulas we have been discussing reflect this.  we mention three:
$$T = (-1/b_0^*)(\sum_{i=0}^n b_i^* \Omega_i^*),$$
$$  \kappa = {{-\tau_0^2} \over {b_0^*}}
{\sum_0^{n-1} b_{i+1}^* E_i^*(\kappa^*)},$$
$$ \kappa_{s^*} = \kappa_s^* \ .$$
We also remark that  the formulas from articles (vii) and (ix)
both have an involutive
nature which also reflects this symmetry.
\newline
\newline
{\bf 5. Evolution equations preserving pseudospherical surfaces}
\medskip
(1) {\it pseudosphericity-preserving deformations and  CTP
vectorfields}:
In a recent paper,
McLachlan and Segur [39]
have investigated differential geometric  aspects of the
evolution of surfaces in $R^3$.
In particular, they  give examples of geometric evolution equations
on surfaces which
preserve the pseudosphericity property, which we call
{\it pseudosphericity-preserving}  evolution equations.
We now describe how their examples fit quite nicely into
the structure described in this paper.

As we have seen, a pseudospherical surface $M$ comes  endowed with a
foliation by curves of constant torsion
(the asymptotic lines).  Thus, a plausible candidate  for a pseudo-sphericity
preserving  vectorfield would be an evolution
equation defined along the asymptotic lines which preserves constant torsion.
  As is shown
in [39], such an evolution equation exists: given $M = M_0$,
 let the asymptotic lines evolve according
to the CTP equation $\gamma_t = Z_1(\gamma)$. Then the resulting surfaces $M_t$
are pseudospherical.

At least for soliton pseudospherical surfaces, this can easily be
extended to {\it any} evolution from
the CTP hierarchy:

{\it Proposition: Let $M = M_0$ be a soliton pseudospherical surface.
  Let the asymptotic lines
evolve  according
to $\gamma_t = Z =  \sum_0^N c_i Z_i$.
  Then the resulting surface $M_t$ at any time $t$ is pseudospherical.
}
\newline
The proof uses the commutativity of the LIH evolution equations.
  Let $\gamma = \gamma(s,0)$
be  an asymptotic
curve for $M$; by definition, $\gamma$ is a planar-like soliton.
  The evolution $\gamma_t =T^*$,
starting with $\gamma$, sweeps out $M$.
  But along $\gamma$, $T^*$ is just a linear combination
of elements from LIH.  This is also true for $Z$.
  {\it All} the evolution equations from LIH
preserve critical points for the functionals  associated  to LIH,
 including $Z$.  Using  the CTP
property of $Z$,  the deformations of $\gamma$ under $Z$ must all be
planar-like solitons of the
same type.
  Commutativity of the $Z$ and $T^*$  evolution equations
implies that any time $t$,
$\gamma(s,t)$ is an asymptotic curve for the surface $M_t$,
 which is therefore pseudospherical.
\newline
(2) {\it pseudo-sphericity preserving
vectorfields of mixed type}:
  In the previous paragraph, the deformations of the
pseudospherical surface $M$ were defined in terms of the
evolution of its asymptotic line foliation. One
could also  define an evolution in terms of  the conjugate line foliation,
as well as evolutions
which combine the two:
$\gamma_t =  Z + Z^* = \sum_0^N c_i Z_i  + \sum_0^N c_i^* Z_i^*$.
 In [39],
McLachlan and Segur essentially ask if evolution equations of this
type are integrable.
Using (ix) from the
previous section, we can answer in the affirmative,
again assuming $M$ is a soliton surface.  The
reasoning is simple:  along such a surface, the $A_i^*$,
 and therefore the $Z_i^*$, can be expressed
in terms of the $A_i$.  In fact, one checks that the $Z_i^*$ are
linear combinations of the $Z_i$,
hence the second summand is redundant.


\medskip
\medskip
{\it Acknowledgements}:  We have already
cited the paper by Melko-Sterling [15],
which provides an alternative approach to studying pseudospherical surfaces.
 It was their
work which suggested to us the connection between
pseudospherical surfaces and LIH;
we refer  the reader to that paper and in
particular its interesting and suggestive
computer graphics of pseudospherical surfaces.

We thank the members of the Symbolic Computation
Group at Drexel University (Char, Johnson, Lakshman, Burke-Perline)
for their
help in steering
the author around various obstacles involved with symbolic computation.
Many of the calculations in this paper were facilitated by the use of
the symbolic manipulation program MAPLE. This paper was completed
during a stay at the Fields Institute for Research in the
Mathematical Sciences,
whose hospitality we are happy to acknowledge.

\end

\end
\newpage


\Refs\nofrills{References}
\widestnumber\key{Ge-Di-long23}



\ref \key{\bf 1} \by  G.K. Batchelor
\book  An introduction to fluid dynamics \publ Cambridge
University Press \publaddr
 New York  \yr 1967
\endref  \ms

\ref \key{\bf 2}  \by  G. L. Lamb, Jr.
\book Elements of Soliton Theory
\publ Wiley
\publaddr New York \yr 1980
\endref \ms


\ref \key{\bf 3} \by R. Ricca
\jour Nature \yr 1991 \vol 352 \page 561
\endref \ms


\ref \key{\bf 4} \by  D. Moore and P. Saffman
\jour Phil. Trans. R. Soc. London
\vol A 272 \yr 1972 \page 403
\endref \ms


\ref \key{\bf 5} \by  R. Klein and A. Majda
\jour Physica D
\yr 1991 \vol 49 \page 323
\endref \ms

\ref \key{\bf 6} \by  J. Marsden and A. Weinstein
\jour Physica 7D \yr1983 \page 305
\endref \ms

\ref \key{\bf 7} \by  H. Hasimoto
\jour  J. Fluid Mech.
\yr1972 \vol 51 \page 477
\endref \ms

\ref \key{\bf 8}
\manyby J. Langer and R. Perline
\jour Appl. Math. Lett.
\yr 1990
\vol 3(2)
\page 61
\endref \ms

\ref \key{\bf 9}
\bysame
\jour J. Nonlinear Sci.
\yr 1991 \vol 1 \page 71
\endref \ms

\ref \key{\bf 10} \by N. Ercolani and R. Montgomery
\jour Phys. Lett. A
\yr 1993 \vol 180 \page 402
\endref \ms

\ref \key{\bf 11} \by H. Dobriner
\jour Acta Math.
\vol 9
\yr 1886 \page 73
\endref \ms


\ref \key{\bf 12} \by A. Enneper
\jour Abh. K\"onigl. Ges. Wissensch. G\"ottingen
\vol 26
\yr 1880
\endref \ms

\ref \key{\bf 13} \by A. B\"acklund
\jour Math.  Ann.
\vol 19
\yr 1882
\page  387
\endref  \ms

\ref \key{\bf 14} \by  A. Sym
\book Geometrical aspects of the Einstein equations
and integrable systems
\publ Lecture Notes in Physics \vol 239 \yr 1985 \page 154
\endref \ms

\ref \key{\bf 15} \by M. Melko and I. Sterling
\jour Ann. of Glob. Anal. and Geom.
\vol 11
\page 65
\yr 1993
\endref  \ms


\ref \key{\bf 16} \by R. Sasaki
\jour Nucl. Phys. B
\yr 1979
\vol 154
\page 343
\endref \ms

\ref \key{\bf 17} \by S.S. Chern and K. Tenenblat
\jour Studies in Applied Mathematics
\vol 74(1) \yr 1986 \page 55
\endref \ms

\ref \key{\bf 18} \by D. Struik
\book Lectures on classical differential geometry
\publ Dover Publ.
\publaddr New York \yr 1988
\endref \ms


\ref \key {\bf 19} \by A. Seeger, H. Donth and A. Kochenforder
\jour Z. Phys.
\yr 1953
\vol 134
\page 173
\endref  \ms


\ref \key {\bf 20} \by J.K. Perring and H.R. Skyrme
\jour Nucl. Phys.
\vol 32
\yr 1962
\page 550
\endref  \ms

\ref \key {\bf 21} \by S.L. McCall and E.L. Hahn
\jour  Phys. Rev. Lett.
\vol 18
\page 908
\yr 1967
\endref   \ms

\ref \key{\bf 22} \by K. Uhlenbeck
\jour J. Diff. Geom.
\yr 1989
\vol 30
\page 1
\endref \ms


\ref \key{\bf 23} \by F. Burstall, D. Ferus, F. Pedit and
U. Pinkall
\jour Ann. of Math.
\vol 138
\yr 1993
\page 173
\endref \ms

\ref \key{\bf 24} \by C. Itzykson and J. Zuber
\book Quantum field theory
\publ McGraw-Hill
\publaddr New York
\yr 1980
\endref \ms

\ref \key{\bf 25} \by M. Crampin and D.J. Saunders
\jour Reports on Mathematical Physics
\vol 23
\yr 1986
\page 327
\endref \ms

\ref \key{\bf 26} \by J. Langer and D. Singer
\paper Lagrangian aspects of the Kirkhhoff elastic rod
\paperinfo Case Western Reserve University preprint
\endref \ms

\ref \key{\bf 27}  \by H. Hasimoto
\jour J. Phys. Soc. Japan
\vol 31(1)
\page 293
\yr 1971
\endref \ms

\ref \key{\bf 28} \by S. Kida
\jour J. Fluid Mech.
\vol 112
\yr 1981
\page 397
\endref \ms

\ref \key{\bf 29} \by R. Ricca
\jour Phys. Fluids A
\vol 4(5)
\page 938
\yr 1992
\endref \ms


\ref \key{\bf 30} \by J. Langer and R. Perline
\paper The filament equation, the Heisenberg model,
and the non-linear Schr\"odinger equation
\paperinfo to appear in the Proceedings
of the Fields Institute Workshop: Mechanics
June 1992
\endref \ms

\ref \key{\bf 31} \by J. Langer and R. Perline
\jour J. Math. Phys.
\vol 35(4)
\page 1732
\endref \ms


\ref \key{\bf 32} \by G.L. Lamb
\jour J. Math. Phys.
\yr 1977
\vol 18
\page 1654
\endref \ms

\ref \key{\bf 33} \by Fukumoto and Miyazaki
\jour J. Fluid. Mech.
\yr 1991
\vol 222
\page 369
\endref \ms

\ref \key{\bf 34} \by J. Langer and R. Perline
\paper The planar filament equation
\paperinfo to appear in the Proceedings
of the Fields Institute Workshop: Mechanics
June 1992
\endref \ms

\ref \key{\bf 35} \by R. Goldstein and D. Petrich
\jour Phys. Rev. Lett.
\yr 1991 \vol 67 \page 3203
\endref \ms

\ref \key{\bf 36} \by K. Nakayama, H. Segur, and M. Wadati
\jour Phys. Rev. Lett.
\vol 69 \yr 1992 \page 2603
\endref \ms

\ref \key{\bf 37} \by R. Goldstein and D. Petrich
\jour Phys. Rev. Lett.
\vol 69(4)
\yr 1993 \page 555
\endref \ms


\ref \key {\bf 38} \by M. Spivak
\book A comprehensive introduction to differential geometry
\publ Publish or Perish
\publaddr Houston
\yr 1979
\endref \ms


\ref \key{\bf 39} \by R. McLachlan and H. Segur
\paper A note on the motion of curves
\paperinfo preprint
\endref \ms


\ref \key{\bf 40} \by S. Novikov
\book Theory of solitons: inverse scattering methods
\publ N.Y. Consultants Bureau
\publaddr New York
\yr 1984
\endref \ms


%
%
%
%
\endRefs
